\documentclass[a4paper]{PoS}

\usepackage{float,picture}
\usepackage{axodraw4j}
\usepackage{booktabs}
\usepackage{multirow}
\usepackage{amsmath,mathtools}
\usepackage{xcolor}
\usepackage{subfig}
\usepackage{graphicx}

\floatstyle{plain}
\newfloat{feyndiagram}{thp}{lop}
\floatname{feyndiagram}{Diagram}

\title{Neutral Kaon mixing beyond the Standard Model}

\ShortTitle{BSM kaon mixing}

\author{\speaker{R. J. Hudspith}\\
	Dept. of Physics and Astronomy,\\
        York University, Toronto, Ontario, M3J 1P3, Canada\\
        E-mail: \email{renwick.james.hudspith@gmail.com}}

\author{N. Garron\\
        Centre for Mathematical Sciences,\\
        School of Computing, Electronics and Mathematics,\\
        Plymouth University, Plymouth, PL4 8AA, United Kingdom\\
        E-mail: \email{nicolas.garron@plymouth.ac.uk}}
        
\author{A. T. Lytle\\
        SUPA, School of Physics and Astronomy,\\
	University of Glasgow, Glasgow, G12 8QQ, United Kingdom\\
        E-mail: \email{andrew.lytle@glasgow.ac.uk}}
        
\author{The RBC and UKQCD collaborations}

\abstract{We compute the hadronic matrix elements of the four-quark operators needed for the study of $K^0-{\bar K^0}$ mixing beyond the Standard Model. We have used $n_f=2+1$ flavours of domain wall fermion (DWF) at two values of the lattice spacing ($a\approx0.08$ and $a\approx0.11\text{fm}$) and with lightest unitary pion mass of $\approx300\,\text{MeV}$. Renormalisation is performed non-perturbatively and the impact of different intermediate momentum schemes is investigated.}

\FullConference{The 33rd International Symposium on Lattice Field Theory\\
				14 -18 July 2015\\
				Kobe International Conference Center, Kobe, Japan*}

\begin{document}

\section{Introduction}

Neutral kaon mixing provides a description of indirect CP violation in the Standard Model (SM), which was discovered in the decay of $K_L\rightarrow \pi\pi$ in 1964 \cite{PhysRevLett.13.138}. In the SM this mixing is mediated by the W-boson and one of the two leading order contributions is given by the box diagram on the left of Diagram.\ref{pic}. In lattice simulations we cannot directly measure the contribution of this diagram but the operator product expansion (OPE) allows us to separate the low and high energy scales and compute the low energy, non-perturbative matrix element from the effective vertex shown on the right.

\begin{feyndiagram}
\begin{center}
\fcolorbox{white}{white}{
  \begin{picture}(400,128) (160,-105)
    \SetWidth{1.0}
    \SetColor{Black}
    \Line[arrow,arrowpos=0.5,arrowlength=5,arrowwidth=2,arrowinset=0.2](176,0)(240,0) 
    \Line[arrow,arrowpos=0.5,arrowlength=5,arrowwidth=2,arrowinset=0.2](240,-96)(176,-96)
    \Photon(240,-96)(240,0){7.5}{5}
    \Line[arrow,arrowpos=0.5,arrowlength=5,arrowwidth=2,arrowinset=0.2](240,0)(320,0)
    \Photon(320,-96)(320,0){7.5}{5}
    \Line[arrow,arrowpos=0.5,arrowlength=5,arrowwidth=2,arrowinset=0.2](320,0)(384,0)
    \Line[arrow,arrowpos=0.5,arrowlength=5,arrowwidth=2,arrowinset=0.2](384,-96)(320,-96)
    \Line[arrow,arrowpos=0.5,arrowlength=5,arrowwidth=2,arrowinset=0.2](320,-96)(240,-96)
    \Text(205,12)[lb]{\Large{\Black{$d$}}}
    \Text(205,-120)[lb]{\Large{\Black{$s$}}}
    \Text(272,12)[lb]{\Large{\Black{$U_1$}}}
    \Text(272,-120)[lb]{\Large{\Black{$U_2$}}}
    \Text(352,-120)[lb]{\Large{\Black{$d$}}}
    \Text(352,12)[lb]{\Large{\Black{$s$}}}
    \Text(201,-52)[lb]{\Large{\Black{$W$}}}
    \Text(346,-52)[lb]{\Large{\Black{$W$}}}
    \Line[arrow,arrowpos=0.5,arrowlength=5,arrowwidth=2,arrowinset=0.2](440,0)(488,-48)
    \Line[arrow,arrowpos=0.5,arrowlength=5,arrowwidth=2,arrowinset=0.2](488,-48)(536,0)
    \Text(434,12)[lb]{\Large{\Black{$d$}}}
    \Text(434,-120)[lb]{\Large{\Black{$s$}}}
    \Text(537,12)[lb]{\Large{\Black{$s$}}}
    \Text(537,-120)[lb]{\Large{\Black{$d$}}}
    \Line[arrow,arrowpos=0.5,arrowlength=5,arrowwidth=2,arrowinset=0.2](488,-48)(440,-96)
    \Line[arrow,arrowpos=0.5,arrowlength=5,arrowwidth=2,arrowinset=0.2](536,-96)(488,-48)
    \GOval(488,-48)(16,16)(0){0.882}
    \Text(483,-55)[lb]{\Large{\Black{$O_1$}}}
  \end{picture}
}
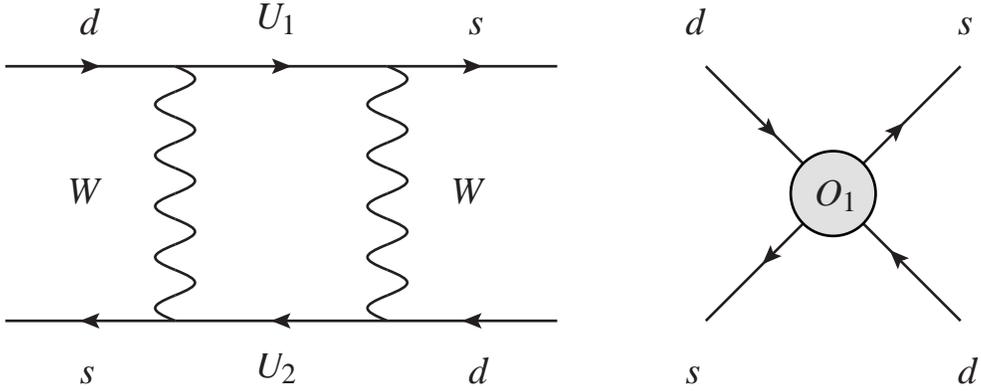
\caption[OPE pinch diagram]{Leading order contribution to neutral kaon mixing in the Standard Model, the right picture illustrates the effective vertex we measure. $U_1,U_2$ can be either $u,c$ or $t$.}\label{pic}
\end{center}
\end{feyndiagram}

In the SM there is only one effective vertex, if we generalise to different mediating particles we can have a larger basis of operators as we have a larger possibility of dirac-color structures. Measuring these new operators can give a model-independent insight into how beyond the Standard Model (BSM) theories could interact with QCD and help constrain the scale of new physics \cite{Bona:2007vi}.

\begin{table}[ht]
\begin{center}
\begin{tabular}{c|ccccc}
\toprule
Collaboration & $B_K$ & $B_2$ & $B_3$ & $B_4$ & $B_5$ \\
\hline
RBC-UKQCD \cite{Boyle:2012qb} & 0.53(2) & 0.43(5) & 0.75(9) & 0.69(7) & 0.47(6) \\
ETM \cite{Bertone:2012cu} & 0.51(2) & 0.47(2) & 0.78(4) & 0.75(3) & 0.60(3) \\
ETM \cite{Carrasco:2015pra} & 0.51(2) & 0.46(3) & 0.79(5) & 0.78(5) & 0.49(4) \\
SWME \cite{Jang:2014aea} & 0.52(2) & 0.53(2) & 0.77(6) & 0.98(6) & 0.75(8) \\
\bottomrule
\end{tabular}
\caption[Previous collabs]{Previous collaboration results for the bag parameters in $\overline{\text{MS}}$ renormalised at $\mu=3\,\text{GeV}$, statistical and systematic errors have been added in quadrature.}\label{tab:prev_evals}
\end{center}
\end{table}
Recently, several groups have measured a dimensionless prescription of these BSM operators (their bag parameters) in modern, dynamical fermion simulations. Their results are shown in Tab.\ref{tab:prev_evals}. There is some tension between these measurements; our previous work \cite{Boyle:2012qb} used a single lattice spacing, $nf=2+1$ flavours and renormalised the operators non-perturbatively using the intermediate, exceptional, RI-MOM scheme. The works \cite{Bertone:2012cu,Carrasco:2015pra} also used the RI-MOM scheme, have $n_f=2$ and $n_f=2+1+1$ flavours respectively and performed an $a^2\rightarrow0$ extrapolation. \cite{Jang:2014aea} have $n_f=2+1$ flavours, renormalised their operators perturbatively and also performed an $a^2\rightarrow0$ extrapolation.

We use the domain wall fermion (DWF) prescription with $n_f=2+1$ dynamical fermion flavours. This prescription has good chiral symmetry properties and leading $\mathcal{O}(a^2)$ scaling. We extend our previous work \cite{Boyle:2012qb} by adding a second lattice spacing to quantify discretisation effects and by using both exceptional and non-exceptional kinematics for our intermediate renormalisation schemes.

\section{Background}

\subsection{Operators}

The $\Delta S=2$ operators we consider are defined in the SUSY basis \cite{Gabbiani:1996hi} ($a$ and $b$ are color indices and Dirac indices have been suppressed),
\newcommand{\opmacro}[2]{\bigl[ #1 \bigr] \, \bigl[ #2 \bigr]}
\begin{equation}\label{eq:operators}
\begin{gathered}
  O_1 = \opmacro{\bar{s}_a\gamma_\mu(1-\gamma_5)d_a}{\bar{s}_b\gamma_\mu(1-\gamma_5)d_b}, \\
  O_{2} = \opmacro{\bar{s}_a(1-\gamma_5)d_a}{\bar{s}_b(1-\gamma_5)d_b},\quad 
  O_{3} = \opmacro{\bar{s}_a(1-\gamma_5)d_b}{\bar{s}_b(1-\gamma_5)d_a}, \\
  O_{4} = \opmacro{\bar{s}_a(1+\gamma_5)d_a}{\bar{s}_b(1+\gamma_5)d_b},\quad 
  O_{5} = \opmacro{\bar{s}_a(1+\gamma_5)d_b}{\bar{s}_b(1+\gamma_5)d_a}. 
\end{gathered}
\end{equation}

\subsection{NPR}

Our operators need to be renormalised, we perform this non-perturbatively in a scheme accessible to the lattice using both exceptional (RI-MOM) and non-exceptional kinematics (RI-SMOM) \cite{Martinelli:1994ty,Aoki:2010pe}. The operators mix multiplicatively under renormalisation, the pattern of which for DWF is that of the continuum i.e. $O_1$ belongs to a $(27,1)$ irreducible representation of $\text{SU}(3)_L\times \text{SU}(3)_R$, whereas $O_2$ and $O_3$ transform like $(6,\bar{6})$ and $O_4$ and $O_5$ like $(8,8)$ \cite{Garron:2012ex}. Schematically, we compute a renormalisation matrix that mixes operators and has renormalisation condition that at the scale $\mu$ our Landau gauge fixed vertex function matches its tree level perturbation theory result,
\begin{equation}
\begin{gathered}
O_i^{\overline{\text{MS}}}(\mu) = C_{ij}^{\overline{\text{MS}}\leftarrow \text{MOM}}(\mu)\left(\lim_{a^2\rightarrow0}\frac{Z_{jk}^{\text{MOM}}(\mu)}{Z_q^2}O_k(a)\right),\\
Z^{\text{MOM}}(\mu) P(\Lambda(p^2))|_{p^2=\mu^2} = \text{tree}.
\end{gathered}
\end{equation}

As discussed in \cite{Lytle:2014tsa} the RI-MOM scheme suffers from pion pole contamination \cite{Giusti:2000jr}. We must subtract these un-physical poles from our renormalisation matrix to obtain physical results, this is a non-trivial procedure and can allow for difficult to quantify systematics. The RI-SMOM scheme does not suffer from this problem.

\subsection{Measurement types}\label{sec::meases}

We intend to measure various dimensionless quantities based upon the effective operators of Eq.\ref{eq:operators}. In \cite{Donini:1999nn} and \cite{Babich:2006bh} ratios were suggested that give directly the BSM to SM contribution at the physical point,
\begin{equation}
 R_i(\mu) = \left[ \frac{f_K^2}{m_K^2}\right]_{\text{Expt.}}\left[ \frac{m_K^2}{f_K^2} \frac{\langle \bar{K}^0| O_i(\mu) | K^0 \rangle }{ \langle \bar{K}^0 | O_1(\mu) | K^0 \rangle }\right]_{\text{Latt.}}.
\end{equation}

Groups also measure the bag parameters, which give the ratio of the matrix element to its vacuum saturation approximation (VSA),
\begin{equation}
B_K(\mu) = \frac{\langle \bar{K}^0 | O_i(\mu) | K^0 \rangle}{\frac{8}{3}m_K^2 f_K^2},\quad B_i(\mu) = \frac{\langle \bar{K}^0 | O_i(\mu) | K^0 \rangle}{N_i m_K^2 f_K^2 \left( \frac{m_K}{m_u(\mu)+m_s(\mu)}\right)^2}.
\end{equation}
With normalisation factors $N_i=\left(-\frac{5}{3},\frac{1}{3},2,\frac{2}{3}\right)$.

In \cite{Becirevic:2004qd} and \cite{Bae:2013tca} combinations of bag parameters were suggested such that the leading chiral logarithms of these quantities in chiral perturbation theory cancel, we call these the golden combinations,
\begin{equation}
\begin{gathered}
 G_{23}(\mu)=\frac{3B_2(\mu)}{5B_2(\mu)-2B_3(\mu)},\quad G_{45}(\mu)=\frac{B_4(\mu)}{B_5(\mu)},\\
 G_{24}(\mu)=B_2(\mu)B_4(\mu),\quad G_{21}(\mu)=\frac{B_2(\mu)}{B_{K}(\mu)}.
\end{gathered}
\end{equation}

The intention of this work is not only to assess the intermediate-scheme dependence but also to measure the various chiral and discretisation effects of these dimensionless quantities.

\section{Methodology}

\begin{table}[ht]
  \centering
  \begin{tabular}{ c | c | c | c }
    \toprule
    Volume & $a^{-1}\:[\text{GeV}]$& $am^{\text{sea}}_{ud} \, (= am^{\text{val}}_{ud})$ & $m_\pi\:[\text{MeV}]$ \\
    \hline
    \multirow{3}{*}{$24^3 \times 64 \times 16$} & 1.785(5) & 0.005, 0.01, 0.02 &340, 430, 560 \\
        \cline{2-4}
    & $am^{\text{sea}}_{s}$ & $am^{\text{val}}_{s}$ & $am^{\text{phys}}_{s}$\\
        \cline{2-4}
    & 0.04 & 0.04, 0.035, 0.03 & 0.03224(18)\\
    \cline{1-4}
     & $a^{-1}\:[\text{GeV}]$& $am^{\text{sea}}_{ud} \, (= am^{\text{val}}_{ud})$ & $m_\pi\:[\text{MeV}]$ \\
    \cline{1-4}
    \multirow{3}{*}{$32^3 \times 64 \times 16$} & 2.383(9) & 0.004, 0.006, 0.008 &300, 360, 410 \\
    \cline{2-4}
    & $am^{\text{sea}}_{s}$ & $am^{\text{val}}_{s}$ & $am^{\text{phys}}_{s}$\\
    \cline{2-4}
    & 0.03 & 0.03, 0.025 & 0.02477(18) \\
    \bottomrule
  \end{tabular}
  \caption{Summary of our lattice ensembles, more details can be found in \cite{Aoki:2010dy}. For the coarse lattice ($a^{-1}=1.785\text{ GeV}$) we use $155,152$ and $146$ measurements for the $am=0.005,0.01$ and $0.02$ ensembles respectively, although in the final analysis the $am=0.02$ ensemble is omitted.
  For the fine lattice we perform $129, 186$ and $208$ measurements for the $am=0.004, 0.006$ and $0.008$ ensembles respectively. The most recent values of $a^{-1}$ and the physical light and strange quarks can be found in \cite{Blum:2014tka}.}\label{tab:lattparam}
\end{table}

We use Coulomb gauge fixed wall sources, the coarse ensemble was fixed to this gauge using the time-slice by time-slice FASD algorithm of \cite{Hudspith:2014oja}, the fine ensemble data was generated as part of the analysis of \cite{Aoki:2010pe}. For the non-perturbative renormalisation we use momentum sources \cite{Gockeler:1998ye} and partially twisted boundary conditions \cite{Martinelli:1994ty}. The ensembles considered in this work are at heavier pion mass than the physical point, we use unitary light valence quarks to extrapolate to the physical pion mass and partially quenched strange quarks to interpolate to the physical strange mass.

\section{Results}

\begin{figure}[ht!]
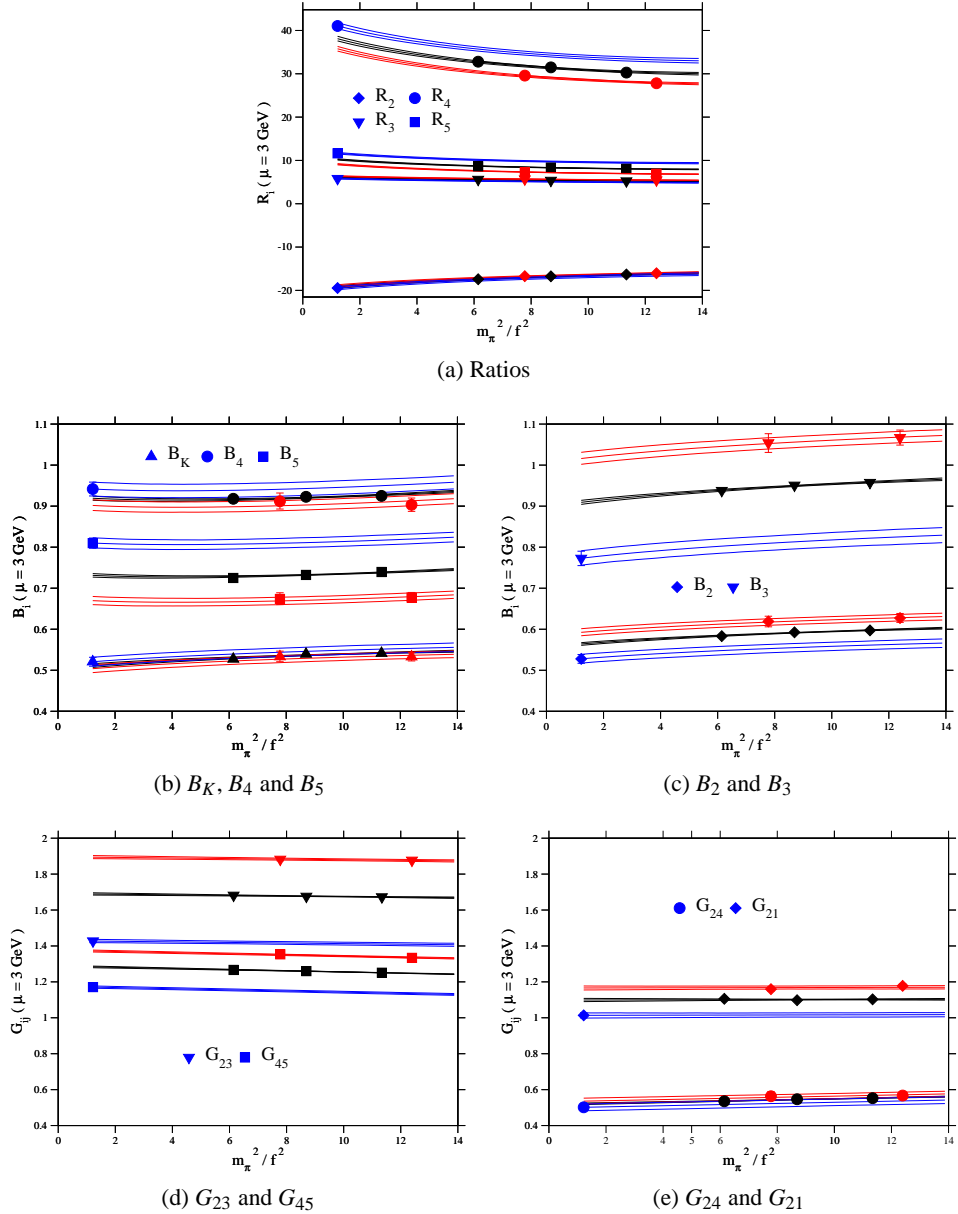

\centering
\subfloat[Ratios]
  {
  \includegraphics[scale=0.23]{chiral_cont_ratios.eps}\label{fig::cc_r}
  }

\subfloat[$B_K$, $B_4$ and $B_5$]
  {
  \includegraphics[scale=0.23]{B145_gg.eps}\label{fig::cc_b145}
  }
\hspace{5pt}
\subfloat[$B_2$ and $B_3$]
  {
  \includegraphics[scale=0.23]{B23_gg.eps}\label{fig::cc_b23}
  }

\subfloat[$G_{23}$ and $G_{45}$]
  {
  \includegraphics[scale=0.23]{Gij_1.eps}\label{fig::cc_G1}
  }
  \hspace{5pt}
\subfloat[$G_{24}$ and $G_{21}$]
  {
  \includegraphics[scale=0.23]{Gij_3.eps}\label{fig::cc_G3}
  }
\caption[]{\textbf{Preliminary} combined chiral and continuum fits for the ratios (\ref{fig::cc_r}), bag parameters (\ref{fig::cc_b145} and \ref{fig::cc_b23}) and the golden parameters (\ref{fig::cc_G1} and \ref{fig::cc_G3}) in the RI-SMOM scheme, renormalised at $3\,\text{GeV}$. The coarse ensemble's data is red, the fine's black. The continuum limit evaluation is blue and the chiral limit is on the left of the figures where the line ends. The combined chiral-continuum result is a filled blue symbol.}\label{fig:chiral_continuum}
\end{figure}

Fig.\ref{fig:chiral_continuum} illustrates the chiral and continuum behaviour of the various quantities from Sec.\ref{sec::meases} renormalised at $\,3\text{ GeV}$ in the RI-SMOM scheme. The coarse ensemble's data and chiral fit is in red, the fine ensemble's is in black and the combined chiral-continuum result is blue with the physical point defined as a filled blue symbol.

For the bag parameters we see reasonably linear behaviour in $m_\pi^2$ as we approach the chiral limit (Figs.\ref{fig::cc_b145} and \ref{fig::cc_b23}) and large discretisation effects upon taking the $a^2\rightarrow0$ limit. For the ratios (Fig.\ref{fig::cc_r}) we see consistent behaviour with our previous work \cite{Boyle:2012qb}, i.e. large ratios of BSM to SM matrix elements. We note that upon taking the $a^2\rightarrow0$ limit these ratios are larger than those of our fine ensemble.

As expected, the approach to the chiral limit for the golden combinations is particularly flat (Figs.\ref{fig::cc_G1} and \ref{fig::cc_G3}) but we do measure large discretisation effects, particularly for the quantity $G_{23}$. More pronounced discretisation effects are measured for these quantities in the RI-MOM scheme but are not shown in Fig.\ref{fig:chiral_continuum}.

\begin{table}[h!]
\begin{center}
\begin{tabular}{c|ccccc}
  \toprule
  Collaboration & $B_K$ & $B_2$ & $B_3$ & $B_4$ & $B_5$ \\
  \hline
  RBC-UKQCD \cite{Boyle:2012qb} & 0.53(2) & 0.43(5) & 0.75(9) & 0.69(7) & 0.47(6) \\
  ETM \cite{Bertone:2012cu} & 0.51(2) & 0.47(2) & 0.78(4) & 0.75(3) & 0.60(3) \\
  ETM \cite{Carrasco:2015pra} & 0.51(2) & 0.46(3) & 0.79(5) & 0.78(5) & 0.49(4) \\
  SWME \cite{Jang:2014aea} & 0.52(2) & 0.53(2) & 0.77(6) & 0.98(6) & 0.75(8) \\
  \hline
  RBC-UKQCD ($\text{RI-MOM}$) & 0.53(1) & 0.42(1) & 0.66(5) & 0.75(3) & 0.56(5) \\
  RBC-UKQCD ($\text{RI-SMOM}$) & 0.53(1) & 0.49(2) & 0.74(7) & 0.92(2) & 0.71(4) \\
  \bottomrule
  \end{tabular}
  \caption{\textbf{Preliminary} results for the bag parameters matched to $\overline{\text{MS}}$ at $\mu=3\,\text{GeV}$ via the intermediate $\text{RI-MOM}$ and $\text{RI-SMOM}$ schemes. Statistical and systematic errors have been added in quadrature, we have not included the perturbative matching systematic in our error. Our collaboration's most up to date and accurate evaluation of $B_K$ should be taken from \cite{Blum:2014tka}.}\label{tab:comparison}
  \end{center}
\end{table}
In Tab.\ref{tab:comparison} we compare the results of this work (the two rows at the bottom of the table) with the results of other collaborations from Tab.\ref{tab:prev_evals}. For the RI-MOM intermediate scheme we are consistent with our previous evaluation at a single lattice spacing and in decent agreement with the most recent ETM evaluation apart from some slight tension in $B_3$. For the RI-SMOM intermediate scheme\footnote{We would like to thank C. Lehner for computing the conversion factors in $\overline{\text{MS}}$ for the $(6,\bar{6})$ operators} we are in considerably better agreement with the most recent results of the SWME collaboration.

\section{Conclusions}

We have attempted to address a tension between different evaluations of the various $\Delta S=2$ matrix elements required for $K^0 -\bar{K}^0$ mixing in and beyond the Standard Model. With the addition of a second lattice spacing we were able to evaluate our discretisation errors and with the use of non-exceptional kinematics in our renormalisation procedure we appear to have an answer as to where this tension originates. It seems that the RI-MOM non-perturbative renormalisation procedure, and the pion pole subtraction this scheme requires, induces a systematic that was previously not well understood.

\section{Acknowledgements}

R. J. H is supported by NSERC of Canada, N. G. is funded by the Leverhulme Trust, research grant RPG-2014-11. A. T. L. is supported by the STFC. The coarse ensemble propagator inversions were performed on the STFC funded DiRAC BG/Q system in the Advanced Computing Facility at the University of Edinburgh.

\bibliographystyle{JHEP}
\bibliography{KKbar2015}

\end{document}